\documentclass[aps,prd,notitlepage,amssymb,amsmath,floatfix,nofootinbib,superscriptaddress]{revtex4-1}

\usepackage{epsfig}
\usepackage{bm}
\usepackage{amssymb}
\usepackage{amsmath}
\usepackage{amsfonts}
\usepackage{color}
\usepackage{slashed}
\usepackage{subfigure}
\usepackage[colorlinks,linkcolor=blue,anchorcolor=black,citecolor=blue]{hyperref}
\allowdisplaybreaks[4]
\usepackage{comment}
\usepackage{tabularx}
\usepackage{hhline}

\begin{document}

\title{Transverse spin polarization as a novel probe of medium-induced transverse-momentum-broadening effect}

\author{Xin-Yu Qin}
\affiliation{SDU-ANU Joint Science College, Shandong University, Weihai, 264209, China}

\author{Yu-Kun Song}
\email{sps\_songyk@ujn.edu.cn}
\affiliation{School of Physics and Technology, University of Jinan, Jinan, Shandong 250022, China}

\author{Shu-Yi Wei}   
\email{shuyi@sdu.edu.cn}
\affiliation{Institute of Frontier and Interdisciplinary Science, Key Laboratory of Particle Physics and Particle Irradiation (MOE), Shandong University, Qingdao, Shandong 266237, China}

\begin{abstract}
The transverse polarization of $\Lambda$ hyperons within unpolarized jets originates from the transverse-momentum-dependent (TMD) fragmentation function $D_{1T}^\perp (z, \bm{p}_\perp, \mu^2)$. In the vacuum environment, the QCD evolution of this TMD fragmentation function is governed by the Collins-Soper equation. However, in the presence of the quark-gluon plasma (QGP) medium, the jet-medium interaction induces a transverse-momentum-broadening effect that modifies the QCD evolution. As a result, the transverse spin polarization of $\Lambda$ hyperons in relativistic heavy-ion collisions differs from that in $pp$ collisions. We demonstrate that this difference serves as a sensitive probe for studying jet-medium interaction, offering a novel perspective through the spin degree of freedom.
\end{abstract}

\maketitle

\section{Introduction}

Fragmentation functions are vital nonperturbative quantities for describing hadron productions in high energy reactions \cite{Metz:2016swz, Boussarie:2023izj, Chen:2023kqw}. In collinear factorization, recent efforts have elevated the quantitative study of the unpolarized fragmentation function $D_{1}(z)$ to a new level of precision \cite{Binnewies:1994ju, Kretzer:2000yf, Bourhis:2000gs, deFlorian:2007aj, Gao:2024nkz, deFlorian:2007ekg, Albino:2008fy, deFlorian:2014xna, deFlorian:2017lwf, Borsa:2021ran, Borsa:2022vvp, Gao:2025hlm,Gao:2025bko}. In the transverse-momentum-dependent (TMD) factorization \cite{Collins:1981uw, Boussarie:2023izj}, the interplay between the transverse momentum and the transverse spin gives birth to numerous interesting phenomena, such as azimuthal asymmetries and emerging hadron polarizations \cite{Bacchetta:2001di, Metz:2002iz, Gamberg:2003eg, Amrath:2005gv, Bacchetta:2007wc, Anselmino:2007fs, Anselmino:2013vqa, Anselmino:2013lza, Wei:2013csa, Wei:2014pma, Anselmino:2015sxa, Kang:2015msa, Anselmino:2015fty, Lu:2015wja, Chen:2016moq, Wei:2016far, Yang:2016mxl, Yang:2017cwi}, providing deep insight into the hadronization mechanism. These effects are modified to a varying extent in relativistic heavy-ion collisions stemming from the jet-medium interaction. The nuclear modification effect thus offers valuable information on the properties of quark-gluon plasma (QGP) matter \cite{Majumder:2004pt, Song:2014sja, Chen:2016vem, Chen:2016cof, Chen:2018fqu}. 

The production of $\Lambda$ hyperons plays an important role in spin physics, since its spin polarization can easily be measured from the anisotropy of its decay products. A variety of models have been proposed to describe its spin contents \cite{Boros:1998kc, Ma:1998pd, Ma:1999gj, Boros:2000ex, Filippone:2001ux}, while only experimental measurements can test them. In recent years, the emerging transverse polarization of $\Lambda$ hyperons within unpolarized jets, driven by the $D_{1T}^\perp (z,\bm{p}_\perp)$ fragmentation function, has attracted significant interest from both the experimental \cite{Belle:2018ttu, Gao:2024dxl} and theoretical communities \cite{Matevosyan:2018jht, Gamberg:2018fwy, Anselmino:2019cqd, Anselmino:2020vlp, DAlesio:2020wjq, Callos:2020qtu, Kang:2020xyq, Boglione:2020cwn, Li:2020oto, Chen:2021hdn, Gamberg:2021iat, Yang:2021zgy, Li:2021txj, Kang:2021ffh, DAlesio:2021pxh,Kang:2021kpt,DAlesio:2021dcx,Chen:2021zrr,Ikarashi:2022yzg,Boglione:2022nzq,DAlesio:2022brl,Zaccheddu:2022qfl,Zhang:2023ugf,DAlesio:2023ozw,DAlesio:2024ope,Gao:2024bfp,Zhao:2024usu}.

The transverse momentum dependence of TMD fragmentation functions is in general governed by the combined effects of nonperturbative contribution and QCD evolution. Although the nonperturbative contribution, so far, can only be extracted from experimental measurements, the QCD evolution for the Sivers function $f_{1T}^\perp (x,\bm{k}_\perp)$, which is the counterpart of the $D_{1T}^\perp (z,\bm{p}_\perp)$ fragmentation function, has been established in Refs.~\cite{Aybat:2011ge, Echevarria:2014xaa}, paving the way for a global analysis of experimental data across different energy scales. Recent works have also predicted the transverse polarization of $\Lambda$-hyperons in semi-inclusive deep-inelastic scatterings (SIDIS) and $pp$ collisions \cite{Kang:2021kpt,Chen:2021zrr,DAlesio:2023ozw,Ji:2023cdh,DAlesio:2024ope,Gao:2024bfp}. In particular, the STAR collaboration at RHIC has recently published preliminary measurements \cite{Gao:2024dxl}. Unlike $e^+e^-$ or SIDIS, TMD factorization in $pp$ collisions is only applicable in the context of the {\it hadron-in-jet} framework \cite{Collins:1993kq, Yuan:2007nd, Kang:2010vd, DAlesio:2010sag, DAlesio:2011kkm, Procura:2011aq, Chien:2015ctp, Bain:2016rrv, Kang:2017glf, DAlesio:2017bvu, Kang:2017btw, Kaufmann:2019ksh, Kang:2020xyq, Kang:2023elg, Bacchetta:2023njc}. Those TMD {\it jet} fragmentation functions, introduced in this scheme, are connected to the TMD {\it parton} fragmentation functions by a matching coefficient, which can be calculated perturbatively \cite{Kang:2017glf}. Since they are identical at the leading order accuracy, we do not distinguish them in the remainder of this paper. The flavor dependence of the polarized fragmentation function $D_{1T}^\perp$ are, so far, poorly known \cite{DAlesio:2020wjq, Callos:2020qtu, Chen:2021hdn}. A global analysis including experimental data from $pp$ and $ep$ collisions can significantly reduce this ambiguity \cite{Chen:2021zrr, Kang:2021kpt, DAlesio:2022brl, DAlesio:2023ozw, DAlesio:2024ope, Gao:2024bfp}.

In relativistic heavy-ion collisions, the formation of a strongly coupled QGP medium introduces additional modifications to jet fragmentation. High energy jets produced in hard scatterings traverse the QGP, undergoing interactions that lead to energy loss and transverse momentum broadening, an effect commonly referred to as jet quenching \cite{Gyulassy:1990ye,Wang:1992qdg, Gyulassy:1993hr,Baier:1994bd,Baier:1996kr,Baier:1996sk,Zakharov:1996fv,Gyulassy:1999zd,Gyulassy:2000fs,Gyulassy:2000er,Wiedemann:2000tf,Guo:2000nz,Wang:2001ifa,Baier:2001yt,Arnold:2002ja,Gyulassy:2003mc,Djordjevic:2003zk,Wicks:2005gt,Djordjevic:2007at,Qin:2007rn,Schenke:2009ik,Jeon:2009yv,Deng:2009ncl,Blaizot:2014bha, Iancu:2014kga, Qin:2015srf, Barata:2020rdn, Barata:2022krd, Ghiglieri:2022gyv, Caucal:2022mpp, Adhya:2022tcn, Li:2023jeh, Mehtar-Tani:2024mvl, Barata:2025htx}. The transverse momentum broadening effect, typically characterized by $\langle \hat q L \rangle$, quantifies the average transverse momentum square acquired by the high energy jet mainly through medium-induced radiation. As a result, the QCD evolution in $pp$ collisions, which occurs in the vacuum, is altered by nuclear effects in relativistic heavy-ion collisions due to interactions with the QGP medium \cite{Iancu:2014kga,Mueller:2016xoc, Mueller:2016gko}. The bottom line is that the medium-induced transverse-momentum-broadening effect introduces additional contribution to the QCD evolution, and therefore modifies the emerging transverse polarization of $\Lambda$ hyperons. 

The QGP is a highly complex system, and numerous theoretical models have been developed to describe both energy loss and transverse momentum broadening effects experienced by energetic partons. A wide range of observables have been proposed and measured in order to build a comprehensive understanding of jet–medium interactions, particular in the context of a realistic QGP medium. The emerging transverse polarization serves as a novel one offering information from the spin degree-of-freedom.

In this paper, we split the QCD evolution of $D_{1T}^\perp$ in relativistic heavy-ion collisions into two parts: the in-medium jet evolution encoding energy loss and transverse momentum broadening effects, and the TMD hadronization in the vacuum governed by the Collins-Soper equation. We explore the nuclear modification on transverse polarization and demonstrate that it serves as a sensitive probe to the jet-medium interaction. The rest of the paper is organized as follows. In Sec. \ref{sec:formalism}, we provide a review of the QCD evolution of $D_{1T}^\perp$ fragmentation function in the vacuum, and present our prescription for incorporating the medium-induced transverse-momentum-broadening effect. In Sec. \ref{sec:numerical}, we present our numerical results for the transverse polarization of $\Lambda$ hyperons in the vacuum environment and estimate the nuclear modification effect. A summary is given in Sec. \ref{sec:conclusion}.

\section{Formalism} 
\label{sec:formalism}

The transverse polarization of $\Lambda$ hyperons within unpolarized jets arises from the TMD fragmentation function $D_{1T}^{\perp} (z, \bm{p}_\perp)$ first proposed in \cite{Mulders:1995dh}. While different notations of $D_{1T}^\perp$ exist in the literature, they all refer to the same transverse polarization–dependent fragmentation function, up to possible sign conventions or normalizations. In this work, we follow the Trento convention \cite{Bacchetta:2004jz}. To be specific, the number density of producing polarized $\Lambda$ hyperons along the $\bm{S}_T$-direction from the unpolarized quark $q$ reads
\begin{align}
{\cal D}_{\Lambda/q} (z, \bm{p}_\perp, \bm{S}_T) = 
\frac{1}{2}\left[
D_{1,\Lambda/q} (z, \bm{p}_\perp) + D^{\perp}_{1T,\Lambda/q}(z, \bm{p}_\perp) \frac{(\hat{\bm{k}} \times \bm{p}_\perp) \cdot \bm{S}_T}{z M_\Lambda} 
\right],
\end{align}
with $M_\Lambda$ the mass of $\Lambda$, $z$ the longitudinal momentum fraction of the fragmenting quark carried by $\Lambda$, and $\bm{p}_\perp$ the transverse momentum of the $\Lambda$ hyperon with respect to the jet axis denoted by $\hat{\bm{k}}$. 

We usually investigated the polarization of $\Lambda$ hyperons along the transverse direction defined by $\bm{n}_T \equiv (\hat{\bm{k}} \times \bm{p}_\perp) / |\bm{p}_\perp|$. Therefore, the number densities for spin up and down (i.e., $\bm{S}_T$ is parallel and antiparallel to $\bm{n}_T$) are given by
\begin{align}
&
{\cal D}_{q\to \Lambda^{\uparrow}} (z, \bm{p}_\perp) = 
\frac{1}{2}\left[
D_{1,\Lambda/q} (z, \bm{p}_\perp) + \frac{|\bm{p}_\perp|}{zM_\Lambda} D^{\perp}_{1T,\Lambda/q}(z, \bm{p}_\perp) 
\right],
\\
&
{\cal D}_{q\to \Lambda^{\downarrow}} (z, \bm{p}_\perp) = 
\frac{1}{2}\left[
D_{1,\Lambda/q} (z, \bm{p}_\perp) - \frac{|\bm{p}_\perp|}{zM_\Lambda} D^{\perp}_{1T,\Lambda/q}(z, \bm{p}_\perp) 
\right].
\end{align}
From the above definition, it is straightforward to obtain the transverse polarization of $\Lambda$ hyperons which reads
\begin{align}
{\cal P}_{T, \Lambda} (z, \bm{p}_\perp) = \frac{{\cal D}_{q\to \Lambda^{\uparrow}} (z, \bm{p}_\perp) - {\cal D}_{q\to \Lambda^{\downarrow}} (z, \bm{p}_\perp)}{{\cal D}_{q\to \Lambda^{\uparrow}} (z, \bm{p}_\perp)+{\cal D}_{q\to \Lambda^{\downarrow}} (z, \bm{p}_\perp)} = \frac{\frac{|\bm{p}_\perp|}{zM_\Lambda} D_{1T,\Lambda/q}^\perp (z, \bm{p}_\perp)}{D_{1,\Lambda/q} (z,\bm{p}_\perp)}. \label{def:pol}
\end{align}

Eq.~(\ref{def:pol}) is fully differential. If we want to study the transverse polarization of $\Lambda$ in a specific phase space, we need to integrate over the phase space separately in the numerator and denominator. For instance, the $\bm{p}_\perp$-integrated transverse polarization of $\Lambda$ hyperons produced from unpolarized jets with a specific quark flavor is given by
\begin{align}
{\cal P}_{T, \Lambda} (z) = \frac{\int d^2 \bm{p}_\perp \frac{|\bm{p}_\perp|}{zM_\Lambda} D_{1T,\Lambda/q}^\perp (z, \bm{p}_\perp)}{\int d^2 \bm{p}_\perp D_{1,\Lambda/q} (z,\bm{p}_\perp)}. \label{def:pol-int}
\end{align}

The factorization scale $\mu$ in TMD fragmentation functions is usually chosen as the energy/transverse momentum of the fragmenting parton $Q$ to minimize higher-order corrections. Their QCD evolution is governed by the Collins-Soper equation \cite{Collins:2011zzd}, which changes in the presence of a QGP medium due to the jet-medium interaction. Since there is an additional power of $|\bm{p}_\perp|$ in the numerator, the nuclear modification factor in the numerator does not cancel with that in the denominator. 

The aim of this paper is to demonstrate that the nuclear modification to the transverse polarization is a sensitive probe to the medium-induced transverse-momentum-broadening effect. Therefore, we adopt a naive model, assuming we have generated a $u$ quark jet with energy $Q$ in $pp$ and $AA$ collisions, and focus on the modification to the QCD evolution due to the medium-induced transverse momentum broadening effect. The Sudakov logarithms and medium-induced transverse momentum broadening effect are quark-flavor independent. Both become larger for the gluon case. Nonetheless, our qualitative conclusion remains the same when summing over all parton flavors.

\subsection{QCD evolution in the vacuum environment}

The QCD evolution of TMD fragmentation functions and parton distribution functions is commonly performed in the impact parameter space by solving the Collins-Soper equation \cite{Collins:2011zzd, Collins:2014jpa, Collins:2016hqq}. The QCD evolution for the Sivers function $f_{1T}^\perp$, which is the partner of $D_{1T}^\perp$ in parton distribution, has been established in \cite{Aybat:2011ge}. Recent papers have also considered the QCD evolution effect in the $D_{1T}^\perp$ fragmentation function \cite{Gamberg:2021iat,DAlesio:2022brl}. In principle, the QCD evolution of the TMD jet fragmentation function differs with that of the TMD parton fragmentation function, since the logarithms in terms of $\ln 1/R^2$ may also need to be resummed at small-$R$ where $R$ is the jet cone-size. The roadmap for this resummation has been presented in~\cite{Kang:2020xyq}. In this work, we adopt the wide jet approximation, which assumes that the jet cone-size is ${\cal O}(1)$. Therefore, logarithms such as $\ln 1/R^2$ becomes very small so that no resummation is required. For completeness, we lay out the essential ingredients of the QCD evolution in the vacuum environment in this subsection. 

Since the TMD evolution is usually performed in the impact parameter space, the TMD fragmentation functions at factorization scale $Q$ are given by Fourier transforms of their impact parameter space counterpart. We obtain
\begin{align}
&
D_{1,\Lambda/q}(z, \bm{p}_\perp, Q) = \int \frac{b_T d b_T}{2\pi} J_0 (|\bm{p}_\perp|b_T/z) \widetilde{D}_{1,\Lambda/q} (z, b_T, Q),
\label{eq:d1-fourier}
\\
&
D_{1T,\Lambda/q}^\perp (z, \bm{p}_\perp, Q) = \frac{M_\Lambda^2}{z|\bm{p}_{\perp}|} \int \frac{b_T^2 db_T}{2\pi} J_1 (|\bm{p}_\perp|b_T/z) \widetilde{D}_{1T,\Lambda/q}^\perp (z, b_T, Q),
\label{eq:d1tperp-fourier}
\end{align}
with $J_{0,1}$ the Bessel functions. $\widetilde{D}_1$ and $\widetilde{D}_{1T}^\perp$ are fragmentation functions in the impact parameter space. The QCD evolution of TMD fragmentation functions in the impact parameter space are governed by the Collins-Soper equation. Utilizing the $b_*$ prescription to separate perturbative and nonperturbative contributions, we arrive at
\begin{align}
&
\widetilde{D}_{1,\Lambda/q}^{\text{vac}}(z, b_T, Q) = \frac{1}{z^2} d_{1,\Lambda/q}(z, \mu_b) \exp\left[ - S_{p} - S_{np} \right],
\label{eq:D1_bT}
\\
&
\widetilde{D}_{1T,\Lambda/q}^{\perp (1),\text{vac}} (z, b_T, Q) = \frac{1}{z^2} \frac{\langle M_D^2 \rangle}{2 M_\Lambda^2} d_{1T,\Lambda/q}^\perp (z, \mu_b) \exp\left[ - S_p - S_{np}^\perp \right],
\end{align}
Here $\mu_b = 2e^{-\gamma_E}/b_*$ and $b_*=b_T/\sqrt{1+b_T^2/b_{\max}^2}$ with $\gamma_E$ the Euler constant and $b_{\max} \approx 1$ GeV$^{-1}$ the infrared cutoff removing nonperturbative contribution denoted by $S_{np}$ from the Sudakov logarithms denoted by $S_p$. $d_{1,\Lambda/q}(z, \mu_b)$ is the collinear unpolarized fragmentation function with factorization scale being set as $\mu_b$. In the numerical evaluation, we adopt the de Florian-Stratmann-Vogelsang (DSV) parameterization \cite{deFlorian:1997zj}. $d_{1T,\Lambda/q}^\perp (z, \mu_b)$ is obtained by fitting the Belle data \cite{Belle:2018ttu} which is related to the unpolarized fragmentation function $d_1$ by
\begin{equation}
d_{1T,\Lambda/q}^\perp (z, \mu_b) = {\cal N}_q (z) d_{1,\Lambda/q}(z, \mu_b),
\end{equation}
with $\mathcal{N}_q(z)$ being parameterized as
\begin{equation}
{\cal N}_q(z) = N_q z^{\alpha_q} (1 - z)^{\beta_q} 
\frac{(\alpha_q + \beta_q - 1)^{\alpha_q + \beta_q - 1}}{(\alpha_q - 1)^{\alpha_q - 1} \beta_q^{\beta_q}}.
\end{equation}
$N_q$, $\alpha_q$ and $\beta_q$ are free parameters taken from the Chen-Liang-Pan-Song-Wei (CLPSW) parametrization \cite{Chen:2021hdn}. The fitting procedure has been laid out in Refs.~\cite{DAlesio:2020wjq, Callos:2020qtu, Chen:2021hdn}. Preliminary results from STAR collaboration \cite{Gao:2024dxl} become also available last year. A comprehensive flavor-separated extraction is not yet feasible due to the lack of sufficient data across different processes. To enable a reliable global analysis, more experimental input from $pp$ and $ep$ collisions will be crucial in the future.

At the next-to-leading logarithmic accuracy, the perturbative Sudakov factor is identical in the evolutions of $\widetilde D_{1}$ and $\widetilde D_{1T}^\perp$, which reads
\begin{align}
S_{p} = \int_{\mu_b^2}^{Q^2} \frac{d\mu^2}{\mu^2} \left[ C_F \frac{\alpha_s (\mu^2)}{2\pi} \ln \frac{Q^2}{\mu^2} - \frac{3}{4} C_F \frac{\alpha_s (\mu^2)}{\pi} \right],
\end{align}
with $C_F=4/3$ the color factor. Notice the above expression is the perturbative Sudakov factor for a quark jet. For a gluon jet, we need to replace $C_F$ with $C_A=3$ and $\frac{3}{4}$ with $\beta_0 = (11-2n_f/3)/12$. Employing the one-loop expression for the running coupling, it is straightforward to obtain the analytic expression for the perturbative Sudakov factor.

The nonperturbative Sudakov factors are usually parameterized as a Gaussian, while the Gaussian widths are different for $D_1$ and $D_{1T}^\perp$. We have
\begin{align}
& 
S_{np} = \frac{\langle p_\perp^2 \rangle}{4} \frac{b_T^2}{z^2}, \label{eq:SNP}
\\
&
S_{np}^\perp = \frac{\langle M_D^2 \rangle}{4} \frac{b_T^2}{z^2}.
\end{align}
Here $\langle p_\perp^2 \rangle = 0.19$ GeV$^2$ and $\langle M_D^2\rangle = 0.118$ GeV$^2$ are Gaussian widths with values taken from Refs.~\cite{Anselmino:2015sxa,Callos:2020qtu}. Due to the positivity constraint, $\langle M_D^2\rangle$ is required to be smaller than $\langle p_\perp^2 \rangle$. Therefore, we cannot use the same nonperturbative Sudakov factor for both unpolarized and polarized TMD fragmentation functions to minimize free parameters. Furthermore, by employing the $b_*$ prescription, we shall also include the $g_2 \ln (b_T/b_*) \ln (Q/Q_0)$ term with $g_2$ and $Q_0$ being free parameters in the nonperturbative Sudakov factor. This term is proposed to address the leftover contribution in replacing $b_T$ with $b_*$ \cite{Sun:2014dqm}. Including this term, we will have four more free parameters to tune with two for each TMD fragmentation function. These parameters can be extracted through a global analysis, when more experimental data at different energy scales become available. We neglect this term in this work and leave it for a future study when more experimental data from RHIC and the LHC become available.

\subsection{QCD evolution in the hot medium environment}
\label{Medium Effects} 

In the relativistic heavy-ion collisions, high energy jets produced from the partonic hard scatterings are accompanied with the strongly coupled QGP, i.e., the hot medium. The jet-medium interaction induces more gluon radiations, which modifies the QCD evolution kernel. The medium-modification effect to the unpolarized fragmentation functions has been investigated in Ref.~\cite{Chen:2016vem}, which allows us to extract the jet transport coefficient $\langle \hat q\rangle$ model-independently. 

\subsubsection{Toy model}

In the toy model, we only focus on the medium-induced transverse-momentum-broadening effect and neglect the energy loss effect which is another face of jet quenching. We follow the same framework and investigate the nuclear modification effect to the transverse polarization of $\Lambda$ in jets. We use $\langle \hat q L \rangle$ to denote the average transverse momentum square gained by the jet through jet-medium interaction. Following the prescription laid out in Refs. \cite{Mueller:2016gko, Mueller:2016xoc, Chen:2016vem}, the QCD evolution of TMD fragmentation functions in the hot medium environment are thus given by
\begin{align}
&
\widetilde{D}^{\text{med}}_{1,\Lambda/q}(z, b_T, Q) = \widetilde{D}_{1,\Lambda/q}^{\text{vac}}(z, b_T, Q) 
\widetilde B (b_T)
, \label{eq:D1_A}
\\
&
\widetilde{D}_{1T,\Lambda/q}^{\perp (1),\text{med}}(z, b_T, Q) = \widetilde{D}_{1T,\Lambda/q}^{\perp (1),\text{vac}}(z, b_T, Q) \, 
\widetilde B (b_T)
. \label{eq:D1Tp_A}
\end{align}
Here $\widetilde B (b_T)$ is the transverse momentum broadening function, which takes the following Gaussian ansatz
\begin{align}
\widetilde B_{\text{G}} (b_T) = \exp\left[ -\frac{1}{4} \langle \hat q L \rangle b_T^2 \right],
\end{align}
with $\langle \hat q L\rangle$ representing the average transverse momentum square gained by the high energy parton while traversing the hot QGP medium. In the event-by-event analysis, the transverse momentum broadening square is given by an integral of jet transportation parameter $\hat q(\bm{x}(t), t)$ along the jet path specified by $\bm{x}(t)$, i.e., $\int_{L} \hat q(\bm{x}(t), t) dt$. The quantity $\langle \hat qL \rangle$ then corresponds to the event-averaged value of this integral, obtained by averaging over initial conditions, QGP fluctuations, and jet paths.

Furthermore, we also explore the effect of non-Gaussian transverse momentum broadening on phenomenological grounds. Inspired by the McLerran–Venugopalan model \cite{McLerran:1993ni, McLerran:1994vd}, we adopt a broadening function that exhibits a power-law tail at large $p_\perp$ \cite{Barata:2020rdn}. The non-Gaussian broadening function is modeled as
\begin{align}
\widetilde B_{\text{nG}} (b_T) = \exp\left[ -\frac{1}{4} \langle \hat q L \rangle b_T^2 \ln\left( e + \frac{2}{\Lambda b_T} \right) \right],
\end{align}
with $\Lambda = 0.2$ GeV. 

By incorporating Eqs.~(\ref{eq:D1_A}-\ref{eq:D1Tp_A}) into those Fourier transforms in Eqs.~(\ref{eq:d1-fourier}-\ref{eq:d1tperp-fourier}), we obtain the medium-modified TMD fragmentation functions. 

In the toy model, we only focus on the medium-induced transverse-momentum-broadening effect and neglect the energy loss effect which is another face of jet quenching. As shown in the next section, incorporating the energy loss effect slightly modifies the overall normalization and has little impact on the transverse momentum distribution.

\subsubsection{Incorporating the energy loss effect}

In this section, we present the formula incorporating the energy loss effect. The whole process is factorized into two stages: (1) The high energy parton undergoes multiple scatterings and parton branchings while traversing the QGP. The emitted partons carry a small fraction of its total energy and induce the transverse momentum broadening effect. (2) The energetic parton hadronizes outside the medium. This process is described by the vacuum transverse momentum dependent fragmentation functions, governed by the standard Collins-Soper equation. We thus obtain
\begin{align}
&
\widetilde D_{1,i}^{\text{med}} (z, b_T, Q, \tau_{\max}) = \sum_j \int_z^1 \frac{d\xi}{\xi} C_{ji} (\xi, \tau_{\max}) \widetilde B_j (b_T) D_{1,j}^{\text{vac}} (\frac{z}{\xi}, b_T, \xi Q),
\\
&
\widetilde D_{1T,i}^{\perp (1), \text{med}} (z, b_T, Q, \tau_{\max}) = \sum_j \int_z^1 \frac{d\xi}{\xi} C_{ji} (\xi, \tau_{\max}) \widetilde B_j (b_T) D_{1T,j}^{\perp (1), \text{vac}} (\frac{z}{\xi}, b_T, \xi Q),
\end{align}
with $\tau_{\max}$ being a dimensionless variable quantifying the per unit ``{\it time}'' \cite{Blaizot:2013hx} of parton-medium interaction. It is also related to the jet quenching parameter $\hat q$, jet path $L$ and jet energy $Q$ \cite{Blaizot:2013hx}. Here, we have assumed the factorization between the energy loss effect and the transverse momentum broadening effect. While the transverse momentum broadening effect is still embedded in the broadening function $\widetilde B$, the energy loss effect is accounted for through the cascade spectrum function $C_{ji}$, which denotes the number density of finding parton $j$ carrying the momentum fraction $\xi$ inside the cascade initialized by parton $i$. Notice that $\langle \hat q L\rangle$ for gluon is about $9/4$ times of that for quark.  

Replacing $C_{ji} (\xi, \tau_{\max})$ with $\delta_{ji} \delta (1-\xi)$, we recover the expression in the toy model, where the energy loss effect is neglected. The parton spectrum $C_{ji} (\xi, \tau_{\max})$ can be obtained by solving the evolution equation developed in \cite{Blaizot:2013hx, Blaizot:2013vha, Mehtar-Tani:2018zba} established on the Baier-Dokshitzer-Mueller-Peigne-Schiff \cite{Baier:1996kr, Baier:1996sk} and Zakharov \cite{Zakharov:1996fv} mechanism. For completeness, we rewrite it in the following compact way
\begin{align}
\frac{\partial C_{ji} (\xi, \tau)}{\partial \tau} = 
&
  \sum_{k=g,q,\bar q}\int_{\xi}^1 dz {\cal K}_{jk} (z) \sqrt{\frac{z}{\xi}} C_{ki} \left(\frac{\xi}{z}, \tau \right) 
- \int_0^1 dz \Bigl\{ {\cal K}_{qq} (z) \delta_{qj} + z [{\cal K}_{gg} (z) + 2n_f {\cal K}_{qg} (z)] \delta_{gj} \Bigr\} \frac{1}{\sqrt{\xi}} C_{ji} (\xi, \tau).
\label{eq:in-medium-evo}
\end{align}
The sum over $k$ runs through all active parton flavors, i.e., $k=g,u,\bar u, d,\bar d, s,\bar s, c,\bar c$ and $n_f = 4$ being the active quark flavor. In the case that the cascade is initialized by parton $i$, the initial condition for the evolution equation reads
\begin{align}
C_{ji} (\xi, \tau = 0) = \delta_{ji} \delta (1-\xi) .
\end{align}
Notice that there is a pole at $z=1$ in the diagonal kernel functions ${\cal K}_{gg}$ and ${\cal K}_{qq}$, which has to be cancelled between the first and the second terms. In the numerical evaluation, we have to combine these two integrands to obtain stable results. The kernel functions \cite{Mehtar-Tani:2018zba} are given by
\begin{align}
&
{\cal K}_{gg} (z) = \frac{2N_c}{2} \frac{(1-z+z^2)^2}{z(1-z)} \sqrt{\frac{N_c(1-z+z^2)}{z(1-z)}},
\\
&
{\cal K}_{qq} (z) = \frac{C_F}{2} \frac{1+z^2}{1-z} \sqrt{\frac{N_c z + C_F (1-z)^2}{z(1-z)}},
\\
&
{\cal K}_{qg} (z) = \frac{T_F}{2} \left[z^2 + (1-z)^2\right] \sqrt{\frac{C_F - N_c z(1-z)}{z(1-z)}},
\\
&
{\cal K}_{gq} (z) = \frac{C_F}{2} \frac{1+(1-z)^2}{z} \sqrt{\frac{N_c(1-z) + C_F z^2}{z(1-z)}},
\end{align}
with $N_c =3$, $C_F = 4/3$ and $T_F=1/2$. Our ${\cal K}_{qg}$ function differs with that in Ref. \cite{Mehtar-Tani:2018zba} by a factor of $2n_f$. This is because we do not regroup those quark/antiquark spectra into singlet and non-singlets and $2n_f$ explicitly appears in Eq.~(\ref{eq:in-medium-evo}).

\section{Numerical Results} \label{sec:numerical}

Solving the QCD evolution equations laid out in the previous section, we obtain the TMD fragmentation functions at high energy scale $Q$ in both vacuum and medium environments. In this section, we first show the nuclear modification effect to the TMD fragmentation functions and then show that to the transverse polarization of $\Lambda$ hyperons within unpolarized jets.

\subsection{Nuclear modification to the TMD fragmentation functions}

Utilizing the prescription presented in Sec.~\ref{Medium Effects}, we can investigate impact of the medium-induced transverse-momentum-broadening effect, characterized by $\langle \hat q L \rangle$, on the TMD fragmentation functions. In this paper, we first present the toy model results to isolate the transverse momentum broadening effect, where the energy loss effect has not been taken into account. We then show numerical results with the energy loss effect incorporated.

Furthermore, in the numerical evaluation, we use $u$ quark jet as an example and set $z=0.3$. The unpolarized collinear $\Lambda$ fragmentation $d_1 (z,\mu_b^2)$ is taken from the DSV parameterization \cite{deFlorian:1997zj}, while that of the $d_{1T}^\perp (z, \mu_b^2)$ is taken from the CLPSW parameterization \cite{Chen:2021hdn}. As for the $d_{1T}^\perp$ function, two other parameterizations exist, namely the D'Alesio-Murgia-Zaccheddu \cite{DAlesio:2020wjq} and the Callos-Kang-Terry \cite{Callos:2020qtu} parameterizations. Both show strong violations of isospin symmetry. After the isospin symmetry issue was first discussed by the CLPSW parameterization \cite{Chen:2021hdn, Chen:2021zrr}, Refs. \cite{DAlesio:2022brl, DAlesio:2023ozw} reanalyzed the Belle data and proposed a renewed isospin-symmetric version. The recent global analysis of the unpolarized fragmentation function \cite{Gao:2025bko} has also examined the isospin symmetry in the production of different hadrons.

\begin{figure}[h!]
\centering
\includegraphics[width=0.32\textwidth]{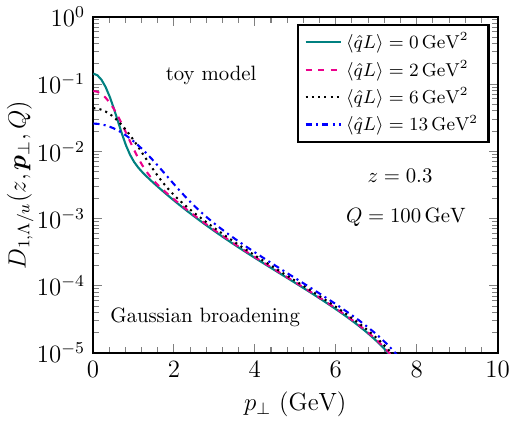}
\includegraphics[width=0.32\textwidth]{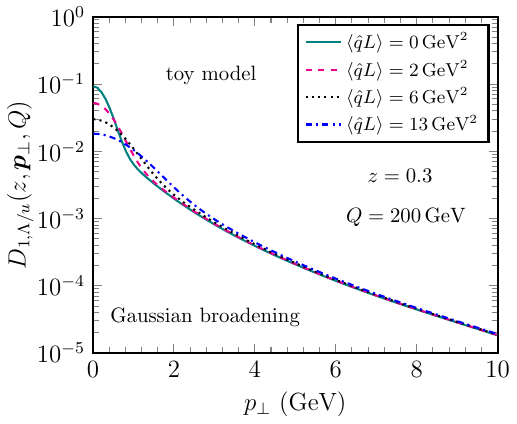}
\includegraphics[width=0.32\textwidth]{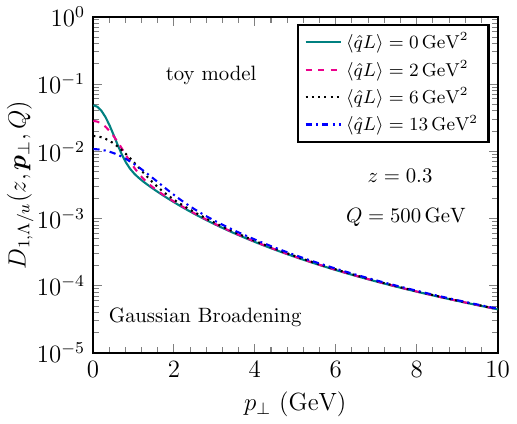}
\\
\includegraphics[width=0.32\textwidth]{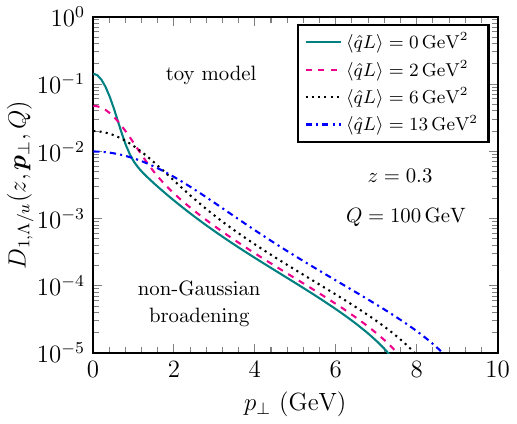}
\includegraphics[width=0.32\textwidth]{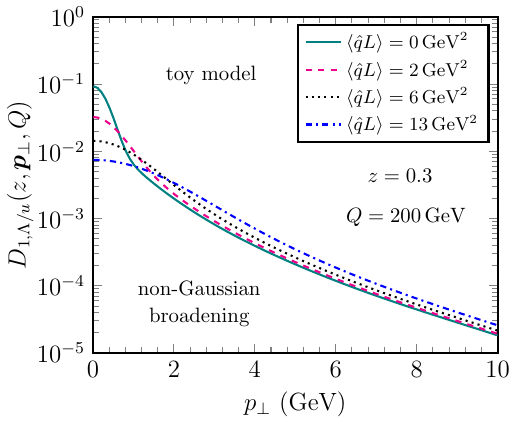}
\includegraphics[width=0.32\textwidth]{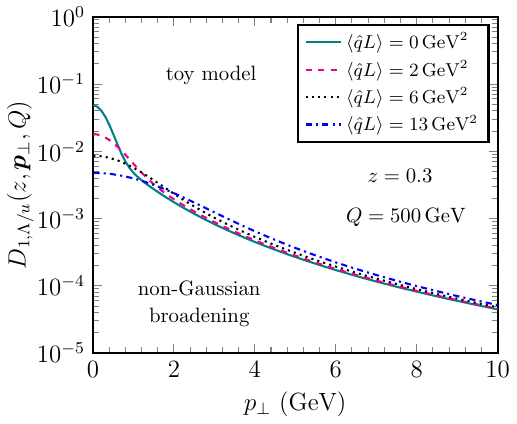}
\caption{Nuclear modification to the unpolarized TMD fragmentation functions $D_{1} (z, \bm{p}_\perp, Q)$ as a function of $|\bm{p}_\perp|$ at $z=0.3$ with several typical energy scales in the toy model. The upper panel is for the Gaussian broadening, and the lower panel is for the non-Gaussian broadening.}
\label{fig:D1}
\end{figure}
\begin{figure}[h!]
\centering
\includegraphics[width=0.32\textwidth]{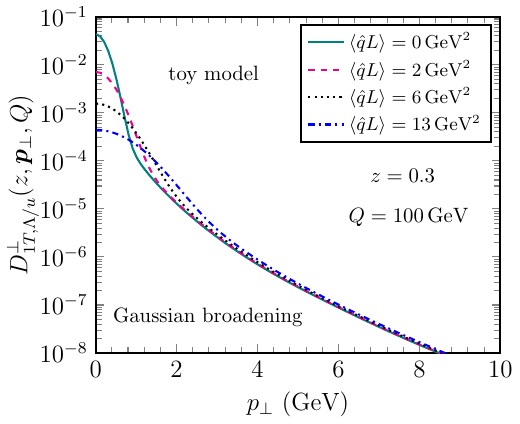}
\includegraphics[width=0.32\textwidth]{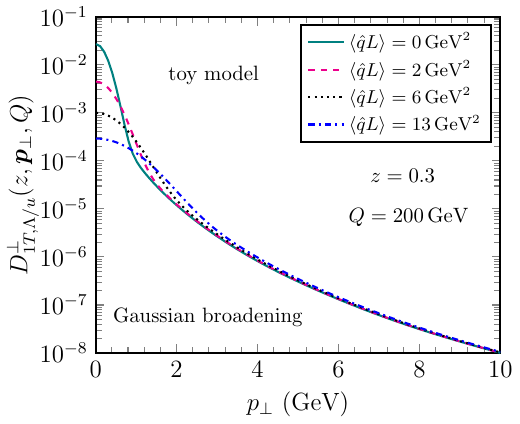}
\includegraphics[width=0.32\textwidth]{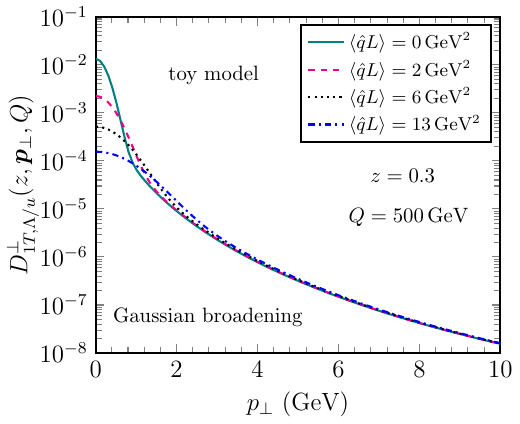}
\\
\includegraphics[width=0.32\textwidth]{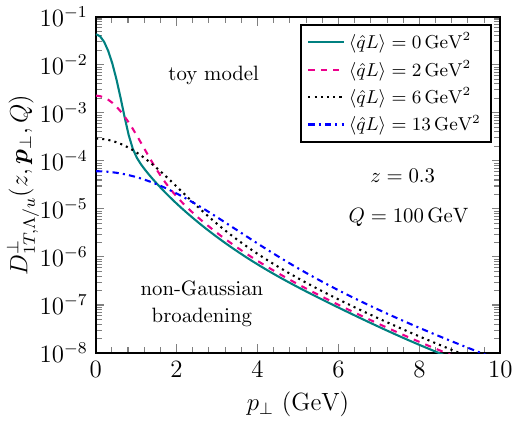}
\includegraphics[width=0.32\textwidth]{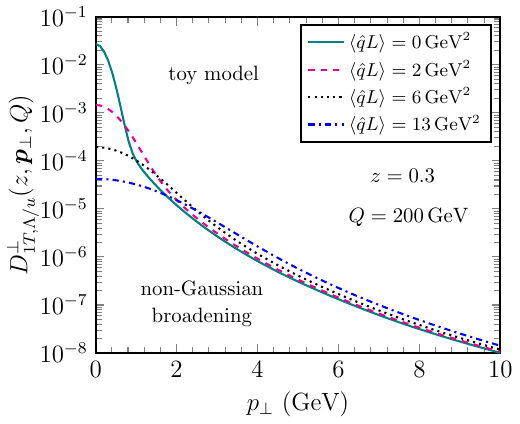}
\includegraphics[width=0.32\textwidth]{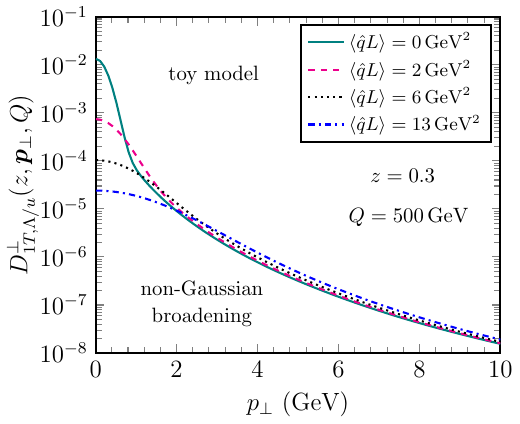}
\caption{Nuclear modification to the polarized TMD fragmentation functions $D_{1T}^\perp (z, \bm{p}_\perp, Q)$ as a function of $|\bm{p}_\perp|$ at $z=0.3$ with several typical energy scales. }
\label{fig:D1Tp}
\end{figure}

We first present those toy model results. The unpolarized TMD fragmentation functions $D_1$ with four typical values of $\langle \hat q L \rangle$ are shown in Fig.~\ref{fig:D1}. While the upper panel is for the Gaussian broadening, the lower panel is for the non-Gaussian case. Generally speaking, as $\langle \hat q L\rangle$ increases, there are more contributions from large $\bm{p}_\perp$ region and less contributions from the small $\bm{p}_\perp$ region. This is an expected feature from the transverse-momentum-broadening effect. In the Gaussian broadening, the modification is more significant at small $\bm{p}_\perp$ and becomes barely visible at large $\bm{p}_\perp$. The reason is given in the following. At small-$\bm{p}_\perp$, the nonperturbative effect is relatively important. The Gaussian broadening resembles an increase of the nonperturbative Gaussian width. Therefore, it plays an important role at very small-$\bm{p}_\perp$. On the other hand, at relatively large $\bm{p}_\perp$ (still much smaller than $Q$), the perturbative Sudakov logarithm dominates and therefore the medium-induced transverse-momentum-broadening effect becomes negligible. On the other hand, the non-Gaussian broadening function provides a power-law contribution at large $p_\perp$, which modifies the fragmentation function at the perturbative region as shown in the lower panel of Fig.~\ref{fig:D1}. Similar feature is also observed in the nuclear modification of the $D_{1T}^\perp$ fragmentation function shown in Fig.~\ref{fig:D1Tp}. 

\begin{figure}[h!]
\centering
\includegraphics[width=0.32\textwidth]{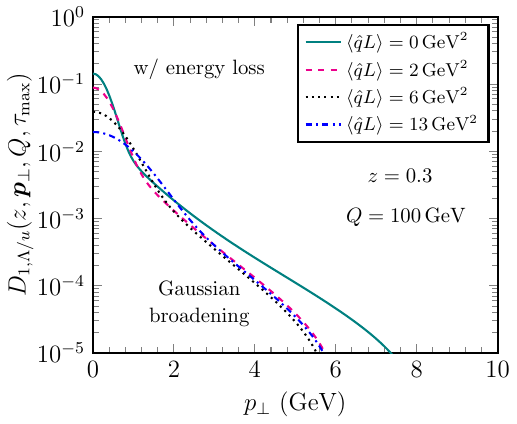}
\includegraphics[width=0.32\textwidth]{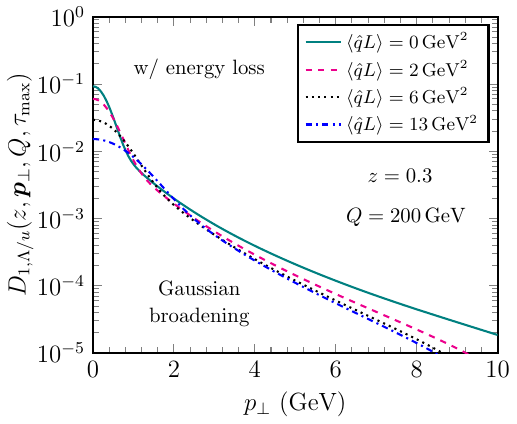}
\includegraphics[width=0.32\textwidth]{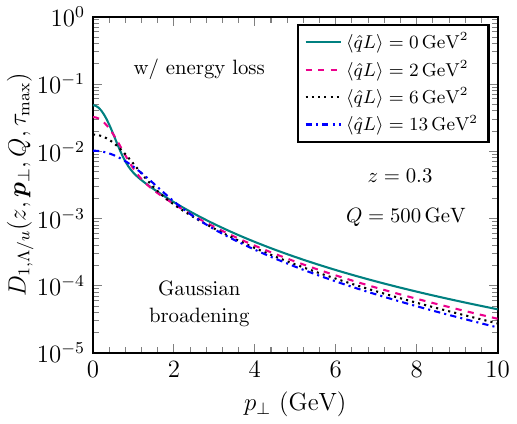}
\\
\includegraphics[width=0.32\textwidth]{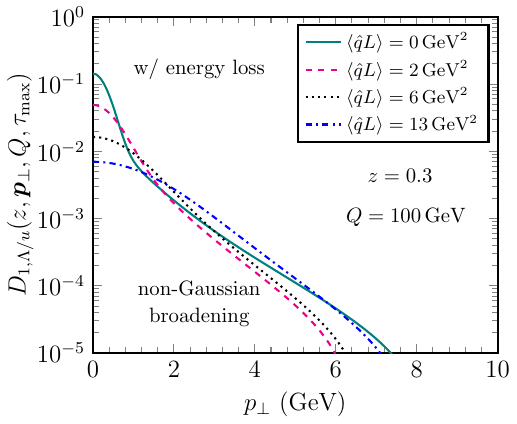}
\includegraphics[width=0.32\textwidth]{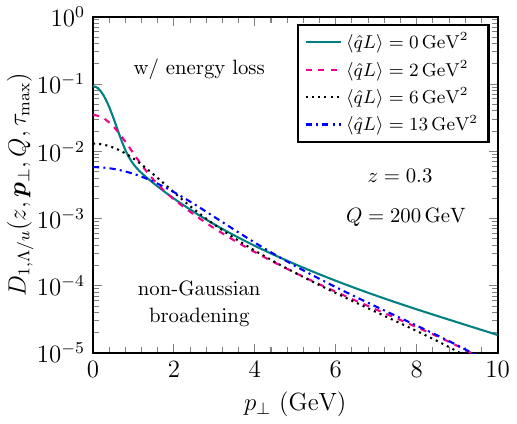}
\includegraphics[width=0.32\textwidth]{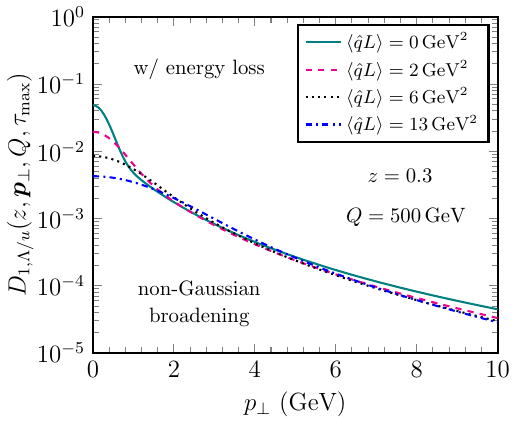}
\caption{Nuclear modification to the unpolarized TMD fragmentation functions $D_{1} (z, \bm{p}_\perp, Q)$ as a function of $|\bm{p}_\perp|$ at $z=0.3$ with several typical energy scales with both transverse momentum broadening and energy loss effects being incorporated. The upper panel is for the Gaussian broadening, and the lower panel is for the non-Gaussian broadening.}
\label{fig:D1_wEL}
\end{figure}
\begin{figure}[h!]
\centering
\includegraphics[width=0.32\textwidth]{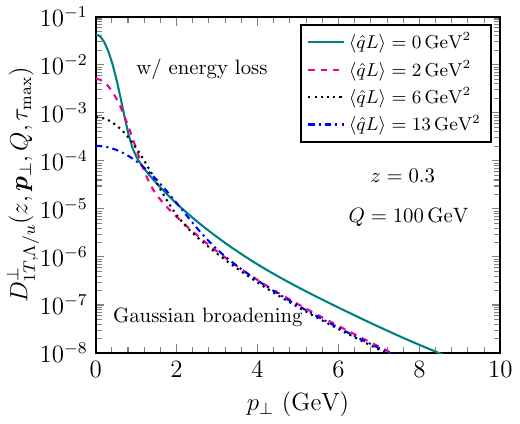}
\includegraphics[width=0.32\textwidth]{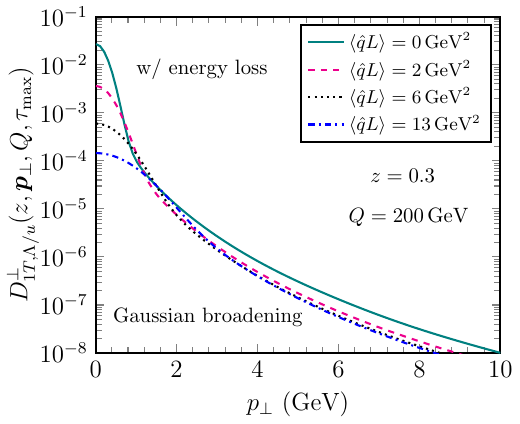}
\includegraphics[width=0.32\textwidth]{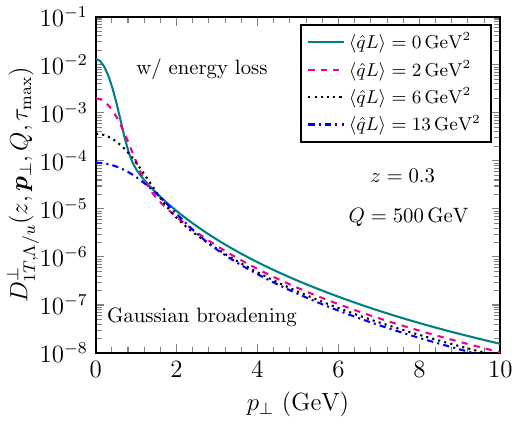}
\\
\includegraphics[width=0.32\textwidth]{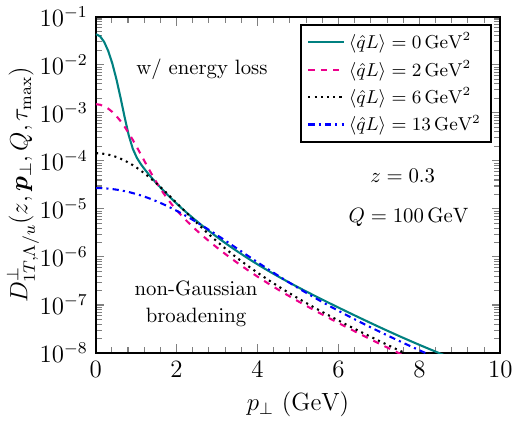}
\includegraphics[width=0.32\textwidth]{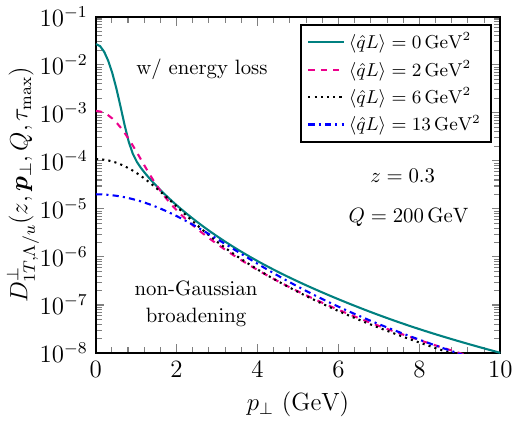}
\includegraphics[width=0.32\textwidth]{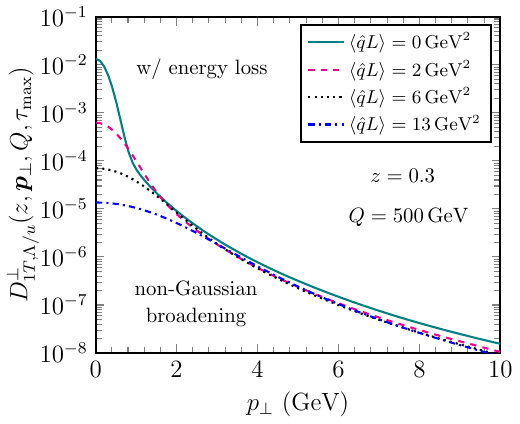}
\caption{Nuclear modification to the polarized TMD fragmentation functions $D_{1T}^\perp (z, \bm{p}_\perp, Q)$ as a function of $|\bm{p}_\perp|$ at $z=0.3$ with several typical energy scales with both transverse momentum broadening and energy loss effects being incorporated. The upper panel is for the Gaussian broadening, and the lower panel is for the non-Gaussian broadening.}
\label{fig:D1Tp_wEL}
\end{figure}

Furthermore, we also incorporate the energy loss effect via the method laid out in Refs. \cite{Blaizot:2013hx, Blaizot:2013vha, Mehtar-Tani:2018zba}. In the numerical calculation, the dimensionless time $\tau_{\max}$ is determined through $\tau_{\max} = \bar\alpha_s \sqrt{\hat qL^2/Q}$ with $\bar\alpha_s=0.2$. In our approximation, we assume that $\langle \hat qL^2\rangle = \langle \hat qL \rangle \langle L\rangle$ with $\langle L \rangle = 4$ fm, which does not hold in a realistic QGP medium. Therefore, $\tau_{\max}$ varies with $\langle\hat qL\rangle$ and $Q$. The numerical results are shown in Fig.~\ref{fig:D1_wEL} for the unpolarized fragmentation function $D_1$ and in Fig.~\ref{fig:D1Tp_wEL} for the polarized fragmentation function $D_{1T}^\perp$. In general, the energy loss effect results in an overalls suppression of of the corresponding fragmentation functions, which suppresses the production rate of single inclusive hadron. 

\subsection{Nuclear modification to the transverse polarization}

\begin{figure}[htbp]
\centering
\includegraphics[width=0.32\textwidth]{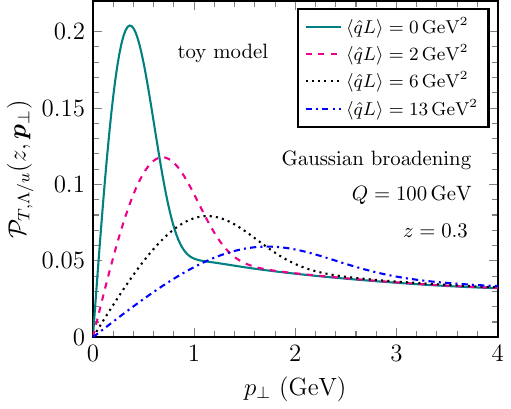} 
\includegraphics[width=0.32\textwidth]{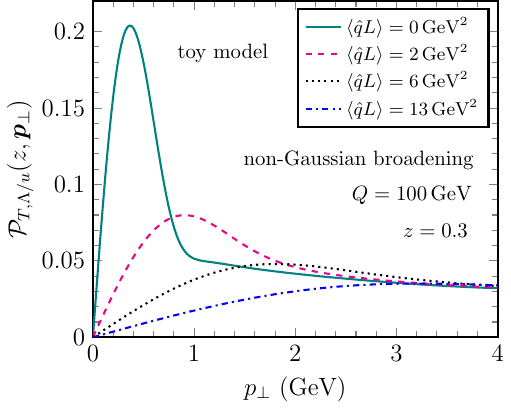} 
\\
\includegraphics[width=0.32\textwidth]{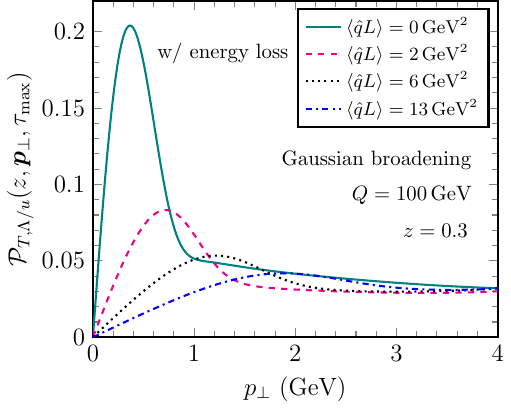} 
\includegraphics[width=0.32\textwidth]{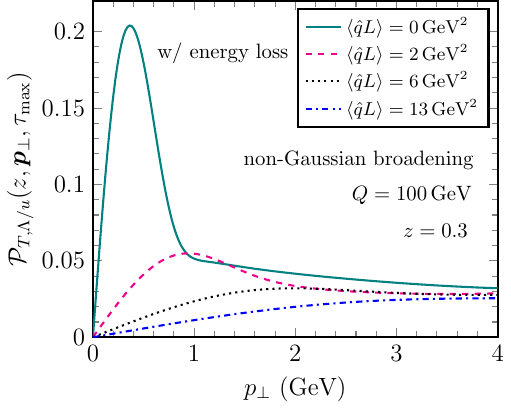} 
\caption{Nuclear modification to the transverse polarization $\mathcal{P}_{T,\Lambda/u}(z, \bm{p}_{\perp})$ of $\Lambda$ hyperons within unpolarized $u$ quark jets as a function of $\bm{p}_\perp$. The upper panel is for the toy model calculation and the lower panel is for that incorporating energy loss effect. The left is for the Gaussian broadening effect and the right is for the non-Gaussian broadening effect with a power-law tail at high-$p_\perp$.}
\label{fig:P}
\end{figure}

Equipped with the TMD fragmentation functions in both vacuum and medium environments, we can make predictions for the nuclear modifications to the transverse polarization of $\Lambda$ hyperons within unpolarized $u$ quark jets.

We first present our predictions for the $\bm{p}_\perp$-differential polarization ${\cal P}_{T,\Lambda/u} (z, \bm{p}_\perp)$ with different $\langle \hat q L \rangle$ values in Fig.~\ref{fig:P}. In the toy model, only the transverse momentum broadening effect is considered. As a result, the transverse polarization of $\Lambda$ hyperons in the medium environment is significantly suppressed at very small-$\bm{p}_\perp$ region. However, at slightly larger $\bm{p}_\perp$, we observe a mild enhancement. This is a universal feature across different energy scales, which occurs because the numerator of ${\cal P}_{T,\Lambda/u} (z, \bm{p}_\perp)$ contains an additional power of $\bm{p}_\perp$. At very large $\bm{p}_\perp$, the perturbative effect dominates, making the medium-induced broadening effect negligible. As shown in Fig.~\ref{fig:P}, the nuclear modification becomes barely visible at $\bm{p}_\perp \sim 4$ GeV. Nonetheless, the nuclear modification to the transverse polarization at small $\bm{p}_\perp$ is very significant, proving a novel tool to explore the jet-medium interaction. 

After incorporating the energy loss effect, we find that the transverse polarization is systematically suppressed as $\langle \hat q L\rangle$ increases. Nevertheless, the transverse momentum distribution continues to exhibit broadening due to medium-induced effects. As shown in the lower panel of Fig.~\ref{fig:P}, the peak of the distribution shifts toward larger $p_\perp$ values with increasing strength of the parton–medium interaction.

\begin{figure}[htbp]
\centering
\includegraphics[width=0.32\textwidth]{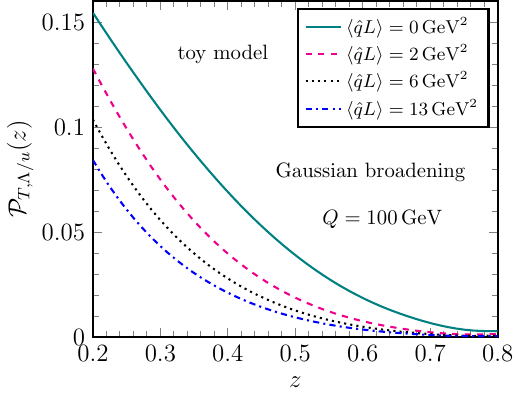} 
\includegraphics[width=0.32\textwidth]{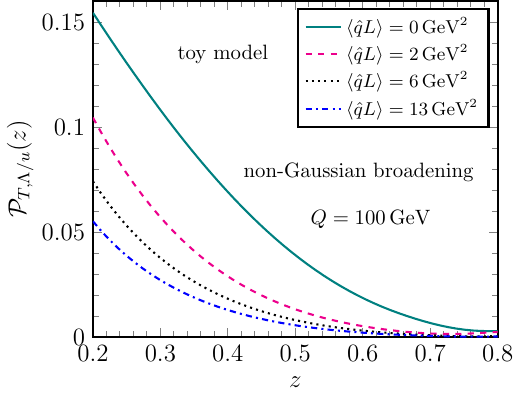} 
\\
\includegraphics[width=0.32\textwidth]{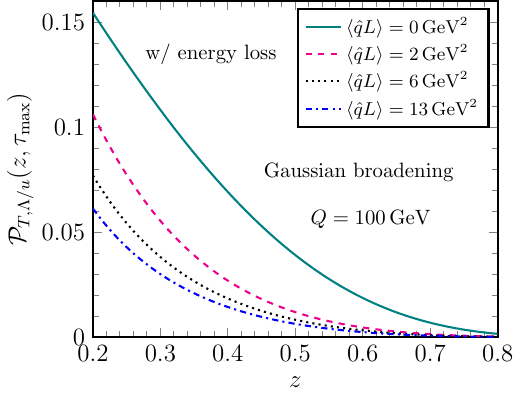} 
\includegraphics[width=0.32\textwidth]{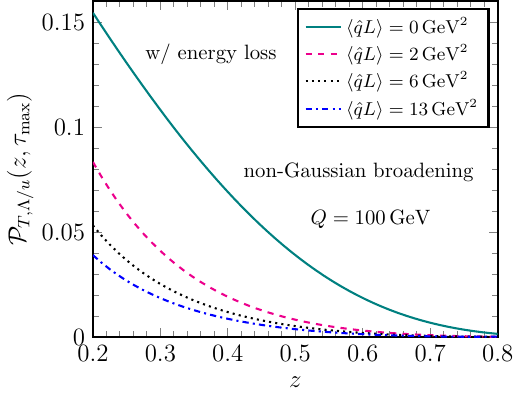} 
\caption{Numerical prediction for the nuclear modification to the $\bm{p}_\perp$ integrated transverse polarization $\mathcal{P}_{T,\Lambda/u}(z)$ of $\Lambda$ hyperons within unpolarized $u$ as a function of the longitudinal momentum fraction $z$. The upper panel is for the toy model calculation and the lower panel is for that incorporating energy loss effect. The left is for the Gaussian broadening effect and the right is for the non-Gaussian broadening effect with a power-law tail at high-$p_\perp$.}
\label{fig:P_int}
\end{figure}

Furthermore, we present our numerical predictions for the nuclear modification of the $\bm{p}_\perp$-integrated transverse polarization in Fig.~\ref{fig:P_int}. The results clearly show that both medium-induced transverse momentum broadening and energy loss effects lead to a suppression of transverse polarization. The behavior of transverse momentum broadening effect can be understood through the positivity constraint \cite{Bacchetta:1999kz, Metz:2016swz}: at the initial scale, positivity is ensured by requiring the Gaussian width of the polarized TMD fragmentation function to be smaller than that of the unpolarized one. Since QCD evolution can only reduce polarization to maintain this constraint at larger factorization scales, the medium-induced broadening effect similarly suppresses polarization to preserve positivity. On the other hand, the impact of energy loss effect reflects the $z$ dependence of $D_{1T}^\perp/D_1$. Therefore, it can in principle increase, as long as it does not exceed one.

\section{Summary} 
\label{sec:conclusion}

In this work, we have investigated the impact of parton-medium interaction on the transverse polarization of $\Lambda$ hyperons within unpolarized jets. By employing the TMD fragmentation formalism, we studied the QCD evolution of the $D_{1T}^\perp$ fragmentation function in both vacuum and QGP environments. Our analysis demonstrates that transverse-momentum broadening effect significantly modifies the differential transverse polarization, leading to suppression effects at low $\bm{p}_\perp$, while a mild enhancement appears at intermediate $\bm{p}_\perp$. At very large $\bm{p}_\perp$, the perturbative contributions dominate, making the medium effects negligible. The energy loss effect, on the other hand, modifies the overall value of the polarization. It does not modify the transverse momentum distribution. Numerical analysis further demonstrates that the $\bm{p}_\perp$-integrated transverse polarization undergoes suppression within the QGP environment, a phenomenon attributable to the influence of positivity constraints.

Our findings underscore that the distinct behavior of transverse polarization, both differential and integrated over $\bm{p}_\perp$, in the QGP medium serves as a sensitive probe of jet-medium interactions. This suggests that future experimental measurements of $\Lambda$ polarization at RHIC and the LHC could provide valuable insights into the properties of the QGP and the mechanisms governing hadronization in heavy-ion collisions. Our findings highlight the potential of transverse polarization as a novel tool for studying jet quenching and the QGP, opening new avenues for both theoretical and experimental exploration.

It would be valuable in future studies to compare the transverse polarization of $\Lambda$ hyperons with that of other hadrons, such as pseudoscalar mesons, vector mesons or other baryons, to help disentangle medium-induced effects from hadronization dynamics. Pseudoscalar mesons, for example, are not subject to spin-dependent fragmentation and can serve as a baseline for isolating purely kinematic broadening effects. In contrast, comparing with other baryons could shed light on the role of quark flavor and spin structure in the polarization mechanism. Moreover, the tensor polarization of vector mesons could provide additional information on spin-dependent aspects of jet quenching. This line of investigation will become increasingly relevant as more experimental data become available.

\section*{Acknowledgments}

We thank Shanshan Cao for helpful discussions. This work is supported by Natural Science Foundation of China under grant No.~12405156 and No.~11505080, the Shandong Province Natural Science Foundation under grant No.~2023HWYQ-011, No.~ZFJH202303, No.~ZR2023MA013 and No.~ZR2018JL006, and the Taishan fellowship of Shandong Province for junior scientists.


\end{document}